# Spectral Interference in High Harmonic Generation from Solids

Yong Woo Kim[1], Tian-Jiao Shao[2,3], Hyunwoong Kim[1], Seunghwoi Han[4], Seungchul Kim[5], Marcelo Ciappina[6], Xue-Bin Bian[2,*], Seung-Woo Kim[1,*]

[1]Department of Mechanical Engineering, Korea Advanced Institute of Science and Technology (KAIST), 291 Daehak-ro, Yuseong-gu, Daejeon 34141, South Korea

[2]State Key Laboratory of Magnetic Resonance and Atomic and Molecular Physics, Wuhan Institute of Physics and Mathematics, Chinese Academy of Sciences, Wuhan 430071, China

[3] School of Physical Sciences, University of Chinese Academy of Sciences, Beijing 100049, China

[4]Institute for the Frontier of Attosecond Science and Technology, CREOL and Department of Physics, University of Central Florida, Orlando, Florida 32816, USA

[5]Department of Optics and Mechatronics Engineering, College of Nanoscience and Nanotechnology, Pusan National University, 2 Busandaehak-ro 63beon-gil, Busan 46241, South Korea

[6]Institute of Physics of the ASCR, ELI-Beamlines project, Na Slovance 2, 182 21 Prague, Czech Republic

*Corresponding authors: swk@kaist.ac.kr, xuebin.bian@wipm.ac.cn

**ABSTRACT:**

**Various interference effects are known to exist in the process of high harmonic generation (HHG) both at the single atom and macroscopic levels**[1–3]**. In particular, the quantum path difference between the long and short trajectories of electron excursion causes the HHG yield to experience interference-based temporal and spectral modulations**[4,5]**. In solids, due to additional phenomena such as multi-band superposition**[6] **and crystal symmetry dependency**[7,8]**, the HHG mechanism appears to be more complicated than in gaseous atoms in identifying accompanying interference phenomena. Here, we first report experimental data showing intensity-dependent spectral modulation and broadening of high harmonics observed from bulk sapphire. Then, by adopting theoretical simulation, the extraordinary observation is interpreted as a result of the quantum path interference between the long and short electron/hole trajectories. Specifically, the**

**long trajectory undergoes an intensity-dependent redshift, which coherently combines with the short trajectory to exhibit spectral splitting in an anomalous way of inverse proportion to the driving laser intensity. This quantum interference may be extended to higher harmonics with increasing the laser intensity, underpinning the potential for precise control of the phase matching and modulation even in the extreme ultraviolet and soft X-ray regime. Further, this approach may act as a novel tool for probing arbitrary crystals so as to adjust the electron dynamics of higher harmonics for attosecond spectroscopy.**

High harmonic generation (HHG) has vigorously been investigated to produce coherent high energy photons in the extreme ultraviolet (EUV) and soft X-ray regime. With much work done on atomic HHG, attention is now being paid to condensed matter[9,10] such as experimental observations of HHG on crystals[11–15] and amorphous materials[16] along with theoretical interpretations on the physical mechanism of solid state HHG[6,17–19]. Intriguing attempts are also being made on surface plasmon-enhanced HHG using nanostructures[20,21] and metasurfaces[22], quantum multi-band interference[6], and symmetry control of harmonic generation[7,8].

The traditional description of HHG in gas phase adopts the semiclassical 3 step model which describes the three steps as ionization, acceleration and recombination[23,24]. Under the influence of a strong driving laser field, the Coulomb potential of the interacting gaseous atom or molecule is distorted to lower the potential barrier, allowing the electron in the ground state to ionize through tunnelling. The ionized electron can be approximated as a free electron, which is accelerated away from the parent ion by the local driving laser field gaining kinetic energy. Finally, the electron oscillates back towards its parent ion as the laser field is reverted, and recombined to emit a high harmonic photon. For each harmonic order below the cutoff, it is possible to determine two unique electron trajectories depending on ionization time, which are said to be long or short depending on their duration of propagation. A similar picture can be given in solid crystals as well, where an electron in a valence band is excited into a conduction band, accelerated within that band and recombined with its parent hole in the original valence band. Unlike in gas HHG, in solids, intraband oscillations during the acceleration step may also generate high harmonics due to the nonlinear gradient of the energy bands[25,26]. Intraband harmonics are known to be synchronized with the driving laser field, having many advantages of generating attosecond pulses from solids[25]. On the other hand, apart from the need for consideration of the band curvature and hole dynamics, the interband mechanism has a similar recollision-like picture as those in gaseous atoms or molecules, thus trajectory analysis can be conveniently applied to explain the resultant harmonic characteristics. Accordingly, in this investigation, we experimentally observe intensity-dependent spectral modulation and broadening of high harmonics and perform the aforementioned analysis to describe the phenomena as due to the quantum path interference of trajectories.

Figure 1 schematically illustrates two trajectories during high harmonics generation eligible for not only gaseous atoms[27] but also solid crystals[28] in the interband picture. The two trajectories are distinguished – short (*s*) and long (*l*) - by the electron's excursion time from birth ($t_b$) till recombination ($t_r$), which can be calculated using the saddle-point equations as described in Supporting Information. As depicted in the reciprocal k-space (Fig. 1b), electron-hole pairs are created through ionization in response to the input electric field, accelerated along the two different trajectories within a band structure, and finally recombined back in the valence band. Each trajectory offers its own distinct quantum path, so the accumulated phase ($\phi_q$) of the *q*-th harmonic turns out larger for the long trajectory owing to its longer excursion time. This trajectory-dependent phase accumulation leads to different frequency chirps, resulting in a slight split in the total sum of the *q*-th harmonic energy (Fig. 1c). In consequence, one harmonic component from the short trajectory interferes with its counterpart from the long trajectory, causing spectral modulations in the resulting harmonic spectrum. An intuitive interpretation is obtained by expressing the accumulated phase $\phi$ and centre frequency $\Omega$ as the functions of the time-varying laser intensity as $\phi_q(t) = -\alpha_q I(t)$ and $\Omega_q(t) = q\omega_0 + \alpha_q \frac{\partial I(t)}{\partial t}$ in which $\alpha_q$ is a constant coefficient depending on the harmonic order $q$ as well as the electron trajectory of short or long[3]. The equations imply that both the accumulated phase and the central frequency shift scale with the constant $\alpha_q$ that is expected to be larger for the long trajectory exhibiting a longer excursion time of the electron-hole pair. In theory, the trajectories may become much more complicated when electrons and holes cross over the first Brillouin zone, inducing multiple channels for harmonic emissions.[29] The driving laser field adopted in our experiment, however, is not strong enough to drive the electrons and holes over the boundaries of the Brillouin zone, thus harmonic emission would be dominated by the two long and short trajectories. In addition, note that although we have illustrated the trajectories to start from the Γ point, electrons may be excited from any point along the Brillouin zone and the resultant harmonics would be generated from a superposition of all possible trajectories.

Figure 2 presents the harmonics spectra obtained from a bulk sapphire target by actual experiment combined with theoretical simulation with varying the input laser field strength. Details of experimental and simulation conditions are given in Methods and Supporting Information. The 2-D plots (Figs. 2a and 2b) show the optical spectra from the 7th (114 nm) to 11th (72.7 nm) harmonics observed for driving laser fields ranging from 0.175 V/Å to 0.226 V/Å. A spectral splitting is clearly seen in the 7th harmonic peak between a nominal peak at 114 nm and a redshifted peak at 118 nm. For relatively lower laser fields, the 7th harmonic exhibits a single peak with negligible splitting (Figs. 2c and 2d). With increasing the laser field, the harmonic peak undergoes spectral broadening with the gradual appearance of a secondary redshifted peak. The original peak shrinks as the secondary peak picks up, and the two peaks are completely reversed for higher laser fields (Figs. 2e and 2f). Based on a two-band model of a sapphire crystal, the computational results (Figs. 2b, 2d and 2f) reproduce the extraordinary growth of the secondary peak progressively until its overtake of the original peak as observed in the experiment. Only a slight discrepancy lies in predicting the relative harmonic yield, of which the cause is reckoned attributable to the imperfection arising in estimating the macroscopic phase matching conditions as well as the multiple bands contribution in the two-band model used in the computation. Propagation effects such as self-phase modulation may also influence the phase of the generated EUV radiation,[30] which are assumed to be small in our experiments because the high-intensity interaction length with the medium is relatively short due to the tight focusing geometry used.

Figure 3 presents theoretical estimations made by numerically solving the semiconductor Bloch equations (SBE) with the application of the saddle point analysis following the computational procedure given in Supporting Information. Specifically, for the bulk sapphire crystal target used in our HHG experiments, a two-band structure model was established using the density functional theory (DFT) with an underestimated band gap adjusted to 9.45 eV (Fig.3a). Other computation parameters were decided in accordance with actual experimental conditions. The harmonic yield (Fig.3b) was first calculated to disclose that the interband harmonics are about 1-3 orders of magnitude higher than the intraband harmonics. This initial finding led us to consider only the interband harmonics in further simulation given hereafter. Then, using the saddle-point semiclassical equations of motion of the electron/hole pair, the harmonic photon energy was obtained as a function of ionization time and re-collision time (Fig. 3c). For a certain ionization time given within a driving laser's optical cycle, there exists a corresponding recombination time at the same energy level as the harmonic photon energy of each trajectory. The electron excursion time for the cutoff harmonic trajectory was determined to be ~0.65 optical cycle (o.c.), which corresponds to 1.74 fs assuming that 1 o.c. equals 2.67 fs. Accordingly, in our computation, ionizations that occur earlier than the cutoff ionization time are treated to recombine later than the cutoff recombination time (long), while ionizations that occur later than the cutoff ionization time recombine earlier than the cutoff recombination time (short). Since the process from ionization to recombination happens within a single optical cycle, trajectories with propagation time less than 1.74 fs are taken as short while trajectories with propagation time between 1.74 fs and 2.67 fs are taken as long. Note that the cutoff harmonic is positioned around the 6.5$^{th}$ harmonic when electrons are excited at the Γ point, and it becomes higher for electrons excited with different crystal momentum values, until the maximum band gap energy is reached. This cutoff energy is obtained from a classical prediction using the field-free band structure, which is expected to extend when accounting for quantum effects in laser-dressed systems.[31,32] Adopting the Floquet picture, the band structures are thought to oscillate in time and the recombination of electrons and holes may occur at extended band gaps relative to the Bloch band structures. On the other hand, due to the flat band structure, electrons excited from many different k points and not just near the Γ point may contribute to the generation of the 7$^{th}$ harmonic which is responsible for the serious modulation of the spectrum.

By adjusting the integration time interval in computation, the long and short trajectory contributions of the 7$^{th}$ harmonic were separated. The results (Figs.3d and 3e) reveal redshifts and blueshifts of the long and short trajectory contributions respectively, for different driving laser intensities. As opposed to the slight blueshift occurring in the harmonics from the short trajectories, the long trajectory harmonics undergo significant amounts of redshift as expected. The directions of frequency shift are determined by the increasing and decreasing gradient of the leading and trailing edges of the driving laser pulse. In theory, the amount of frequency shift should increase with increasing intensity, however in our case the frequency shift remains more or less constant for varying intensities, as we assume an ideal Gaussian driving pulse in our simulations and the redshift due to the intensity gradient decrease in the trailing edge is compensated by the blueshift in the leading edge. Finally, when the two trajectories are coherently combined (Fig. 3f), the resulting spectra clearly show the spectral broadening, secondary peak growth, and peak reversal. This provides a solid evidence for the quantum path interference observed in our experiments. It is also worthwhile to note that the theoretical predictions described so far is not limited to a specific band structure or harmonic order, thereby being applicable to other materials and harmonic orders (See Supporting Information).

Figure 4 provides additional simulation result on how the 7$^{th}$ harmonic experiences spectral splitting and frequency modulation by the quantum interference between the short and long trajectories. The real (Figs. 4a-4c) and imaginary (Figs. 4d-4f) parts of the interband polarization calculated for the long, short and coherently added contributions are shown in the frequency domain together with corresponding harmonic spectra (Figs. 4g-

4h). The result signifies that even though the amount of redshift induced in either the short or long trajectory is not substantial, the combined harmonic splits into two peaks together with a significant spectral modulation. Figure 5 shows further analysis of the effect of the dephasing time on the quantum interference in terms of the harmonic spectrum and spectrogram. The results are first calculated for a dephasing time of 2.67 fs, corresponding to 1 o.c., in which the long dephasing time permits the coherence between the electron and hole to be maintained for both the short and long trajectories along with a high level of spectral splitting (Fig. 5a and 5b). When the dephasing time is reduced to 0.668 fs (0.25 o.c.), shorter than the excursion time of the cutoff trajectory of 1.74 fs (0.65 o.c.), the long trajectory is suppressed due to the loss of coherence between the electron and hole, leaving only the short trajectory component (Figs. 5c and 5d). This result verifies the role of the long trajectory in the consequence of quantum interference.

To conclude, the anomalous phenomenon of spectral modulation and broadening observed in our experiments of high harmonic generation from bulk sapphire has been clarified as a consequence of the quantum interference between the long and short trajectories. Specifically, the intensity-dependent phase delay induced in high harmonics causes frequency redshifting particularly in the long trajectory in line with the harmonic yield dominance along the trailing edge of the driving laser. The redshifted long trajectory coherently interferes with the short trajectory of less intensity-dependency, exhibiting extraordinary features such as the broadening and splitting of harmonic peaks followed by complete peak reversal. Our theoretical simulation based on a two-band solid model excellently reproduces all the experimentally observed features in evidence of the dominance of interband harmonics on the coherent superposition between the short and long trajectories. These findings lead to the possibility that the individual trajectory phases may be controlled to an optimal condition so as to achieve a designated spectrum with sufficient yield in solid harmonics. Further investigations will open the potential for direct modifications of the electron dynamics within the crystal band structure for the purpose of flexible selections of high harmonics with subsequent yield enhancement in the EUV and soft X-ray regime, offering new opportunities in advancing high harmonic spectroscopy.

**Methods**

**Sapphire target for HHG experiment.** The apparatus used to obtain the experimental results of Fig. 2 is given schematically in Supporting Information. The sapphire target was cut along the A-plane to a 430 µm thickness and then polished to a fine surface roughness of < 0.3 nm. The driving laser was aligned with its polarization direction fixed along the Γ-A direction of the sapphire target. The band structure and transition dipoles of the sapphire target were estimated from the α-Al$_2$O$_3$ data given in the Vienna Ab initio Simulation Package (VASP). The dynamics of electron-hole pairs in the laser field was calculated by solving the semiconductor Bloch equation. The saddle-point equations were used to determine the ionization and recombination time of the long and short trajectories as described in Supporting Information.

**Laser illumination for HHG experiment.** The driving laser was produced from a Ti:Sapphire oscillator emitting short pulses of 12 fs duration at a 75 MHz repetition rate with a 375 THz centre frequency. The incident laser was focused using an achromatic lens onto the rear surface of the sapphire target with a spot diameter of 5 µm. Assuming a Gaussian pulse, the peak intensity in vacuum is calculated as $I = 1.88 \times P_a/f_r\tau A$ with $P_a$ being the average power, $f_r$ the repetition rate, $\tau$ the pulse duration, and $A$ the focal area. Then, in consideration of the reflection loss from the boundary between the sapphire target and vacuum, the actual peak intensity was estimated as $I_{saph} = I \cdot 4n/(n+1)^2$ with the refractive index $n$ being set at 1.76 for sapphire (α-Al$_2$O$_3$)[33]. During the experiment, the average power was adjusted from 210 to 350 mW in 10 mW steps using a pair of half and quarter waveplates, varying the actual peak intensity from 0.72 to 1.2 TW/cm$^2$.

**Spectrum Measurement.** The spectrometer used in the experiment of Fig. 2 was a Rowland-circle grazing incidence type (McPherson 248/301), capable of detecting photons of wavelengths shorter than 310 nm with a spectral resolution of 0.16 nm. Using a toroidal mirror accepting light within a 5° horizontal angle and 8° vertical angle, generated harmonics were refocused through a rectangular slit on a reflection type curved grating of a 998 mm radius of curvature to produce a linear spectrum. For quantitative spectral analysis, a detector unit composed of a microchannel plate (XUV-2040, Brightview Inc.) and a charge coupled device camera (DH420A-FO-195, Andor) was used. The whole spectrometer system was kept under a vacuum condition of $10^{-6}$ Torr.

**Supporting Information**

The Supporting Information is available free of charge available free of charge on the ACS Publications website at DOI:

(1) Illustration of the experimental setup in Figure S1; (2) Theoretical details on the semiconductor Bloch equations and first principle calculations; (3) Numerical calculations showing spectral interference in the

9th harmonic spectra

**ACKNOWLEDGEMENTS**

This work was supported by the National Research Foundation of the Republic of Korea (NRF-2012R1A3A1050386). T.-J.S and X.-B.B were supported by the National Natural Science Foundation of China (NSFC) (Grants No. 91850121, and No. 11674363). M. C. is supported by the project Advanced research using high intensity laser produced photons and particles (CZ.02.1.01/0.0/0.0/16_019/0000789) from European Regional Development Fund (ADONIS). S. K. is supported by the National Research Foundation of the Republic of Korea (NRF-2017R1C1B2006137).

**AUTHOR CONTRIBUTIONS**

The project was planned and overseen by S.-W.K. in collaborations with X.-B.B., S.K. and M.C. Experiments were performed by Y.W.K., H.K. and S.H., and simulation was conducted by T.-J.S. and X.-B.B. All authors contributed to the manuscript preparation.

**References**

(1) Kanai, T.; Minemoto, S.; Sakai, H. Quantum Interference during High-Order Harmonic Generation from Aligned Molecules. *Nature* **2005**, *435*, 470–474. https://doi.org/10.1038/nature03577.
(2) Zaïr, A.; Holler, M.; Guandalini, A.; Schapper, F.; Biegert, J.; Gallmann, L.; Keller, U.; Wyatt, A. S.; Monmayrant, A.; Walmsley, I. A.; et al. Quantum Path Interferences in High-Order Harmonic Generation. *Phys. Rev. Lett.* **2008**, *100*, 143902. https://doi.org/10.1103/PhysRevLett.100.143902.
(3) Heyl, C. M.; Güdde, J.; Höfer, U.; L'Huillier, A. Spectrally Resolved Maker Fringes in High-Order Harmonic Generation. *Phys. Rev. Lett.* **2011**, *107*, 033903. https://doi.org/10.1103/PhysRevLett.107.033903.
(4) He, L.; Lan, P.; Zhang, Q.; Zhai, C.; Wang, F.; Shi, W.; Lu, P. Spectrally Resolved Spatiotemporal Features of Quantum Paths in High-Order-Harmonic Generation. *Phys. Rev. A* **2015**, *92*, 043403. https://doi.org/10.1103/PhysRevA.92.043403.
(5) Nefedova, V. E.; Ciappina, M. F.; Finke, O.; Albrecht, M.; Vábek, J.; Kozlová, M.; Suárez, N.; Pisanty, E.; Lewenstein, M.; Nejdl, J. Determination of the Spectral Variation Origin in High-Order Harmonic Generation in Noble Gases. *Phys. Rev. A* **2018**, *98*, 033414. https://doi.org/10.1103/PhysRevA.98.033414.
(6) Hohenleutner, M.; Langer, F.; Schubert, O.; Knorr, M.; Huttner, U.; Koch, S. W.; Kira, M.; Huber, R. Real-Time Observation of Interfering Crystal Electrons in High-Harmonic Generation. *Nature* **2015**, *523*, 572–575. https://doi.org/10.1038/nature14652.
(7) Langer, F.; Hohenleutner, M.; Huttner, U.; Koch, S. W.; Kira, M.; Huber, R. Symmetry-Controlled Temporal Structure of High-Harmonic Carrier Fields from a Bulk Crystal. *Nat. Photonics* **2017**, *11*, 227–231. https://doi.org/10.1038/nphoton.2017.29.
(8) Liu, H.; Li, Y.; You, Y. S.; Ghimire, S.; Heinz, T. F.; Reis, D. A. High-Harmonic Generation from an Atomically Thin Semiconductor. *Nat. Phys.* **2017**, *13*, 262–265. https://doi.org/10.1038/nphys3946.
(9) Ghimire, S.; Dichiara, A. D.; Sistrunk, E.; Agostini, P.; Dimauro, L. F.; Reis, D. A. Observation of

High-Order Harmonic Generation in a Bulk Crystal. *Nat. Phys.* **2010**, *7*, 138–141. https://doi.org/10.1038/nphys1847.
(10) Luu, T. T.; Garg, M.; Moulet, A.; Goulielmakis, E.; Kruchinin, S. Y.; Moulet, A.; Hassan, M. T.; Goulielmakis, E. Extreme Ultraviolet High-Harmonic Spectroscopy of Solids. *Nature* **2015**, *521*, 498–502. https://doi.org/10.1038/nature14456.
(11) Schubert, O.; Hohenleutner, M.; Langer, F.; Urbanek, B.; Lange, C.; Huttner, U.; Golde, D.; Meier, T.; Kira, M.; Koch, S. W.; et al. Sub-Cycle Control of Terahertz High-Harmonic Generation by Dynamical Bloch Oscillations. *Nat. Photonics* **2014**, *8*, 119–123. https://doi.org/10.1038/nphoton.2013.349.
(12) Vampa, G.; Hammond, T. J.; Thire, N.; Schmidt, B. E.; Legare, F.; McDonald, C. R.; Brabec, T.; Corkum, P. B. Linking High Harmonics from Gases and Solids. *Nature* **2015**, *522*, 462–464. https://doi.org/10.1038/nature14517.
(13) Kim, H.; Han, S.; Kim, Y. W.; Kim, S.; Kim, S. Generation of Coherent Extreme-Ultraviolet Radiation from Bulk Sapphire Crystal. *ACS Photonics* **2017**, *4*, 1627–1632. https://doi.org/10.1021/acsphotonics.7b00350.
(14) You, Y. S.; Wu, M.; Yin, Y.; Chew, A.; Ren, X.; Gholam-Mirzaei, S.; Browne, D. A.; Chini, M.; Chang, Z.; Schafer, K. J.; et al. Laser Waveform Control of Extreme Ultraviolet High Harmonics from Solids. *Opt. Lett.* **2017**, *42*, 1816–1819. https://doi.org/10.1364/OL.42.001816.
(15) Ndabashimiye, G.; Ghimire, S.; Wu, M.; Browne, D. A.; Schafer, K. J.; Gaarde, M. B.; Reis, D. A. Solid-State Harmonics beyond the Atomic Limit. *Nature* **2016**, *534*, 520–523. https://doi.org/10.1038/nature17660.
(16) You, Y. S.; Yin, Y.; Wu, Y.; Chew, A.; Ren, X.; Zhuang, F.; Gholam-Mirzaei, S.; Chini, M.; Chang, Z.; Ghimire, S. High-Harmonic Generation in Amorphous Solids. *Nat. Commun.* **2017**, *8*, 724. https://doi.org/10.1038/s41467-017-00989-4.
(17) Vampa, G.; McDonald, C. R.; Orlando, G.; Klug, D. D.; Corkum, P. B.; Brabec, T. Theoretical Analysis of High-Harmonic Generation in Solids. *Phys. Rev. Lett.* **2014**, *113*, 073901. https://doi.org/10.1103/PhysRevLett.113.073901.
(18) Luu, T. T.; Wörner, H. J. High-Order Harmonic Generation in Solids: A Unifying Approach. *Phys. Rev. B* **2016**, *94*, 115164. https://doi.org/10.1103/PhysRevB.94.115164.
(19) Osika, E. N.; Chacón, A.; Ortmann, L.; Suárez, N.; Pérez-Hernández, J. A.; Szafran, B.; Ciappina, M. F.; Sols, F.; Landsman, A. S.; Lewenstein, M. Wannier-Bloch Approach to Localization in High-Harmonics Generation in Solids. *Phys. Rev. X* **2017**, *7*, 021017. https://doi.org/10.1103/PhysRevX.7.021017.
(20) Han, S.; Kim, H.; Kim, Y. W.; Kim, Y.-J.; Kim, S.; Park, I.-Y.; Kim, S.-W. High-Harmonic Generation by Field Enhanced Femtosecond Pulses in Metal-Sapphire Nanostructure. *Nat. Commun.* **2016**, *7*, 13105. https://doi.org/10.1038/ncomms13105.
(21) Vampa, G.; Ghamsari, B. G.; Siadat Mousavi, S.; Hammond, T. J.; Olivieri, A.; Lisicka-Skrek, E.; Naumov, A. Y.; Villeneuve, D. M.; Staudte, A.; Berini, P.; et al. Plasmon-Enhanced High-Harmonic Generation from Silicon. *Nat. Phys.* **2017**, *13*, 659–662. https://doi.org/10.1038/nphys4087.
(22) Liu, H.; Guo, C.; Vampa, G.; Zhang, J. L.; Sarmiento, T.; Xiao, M.; Buscbaum, P. H.; Vučković, J.; Fan, S.; Reis, D. A. Enhanced High-Harmonic Generation from an All-Dielectric Metasurface. *Nat. Phys.* **2018**, *14*, 1006–1010. https://doi.org/10.1038/s41567-018-0233-6.
(23) Corkum, P. B. Plasma Perspective on Strong Field Multiphoton Ionization. *Phys. Rev. Lett.* **1993**, *71*, 1994–1997. https://doi.org/10.1103/PhysRevLett.71.1994.
(24) Lewenstein, M.; Balcou, P.; Ivanov, M. Y.; L'Huillier, A.; Corkum, P. B. Theory of High-Harmonic Generation by Low-Frequency Laser Fields. *Phys. Rev. A* **1994**, *49*, 2117–2132. https://doi.org/10.1103/PhysRevA.49.2117.
(25) Garg, M.; Zhan, M.; Luu, T. T.; Lakhotia, H.; Klostermann, T.; Guggenmos, A.; Goulielmakis, E. Multi-Petahertz Electronic Metrology. *Nature* **2016**, *538*, 359–363. https://doi.org/10.1038/nature19821.
(26) Garg, M.; Kim, H. Y.; Goulielmakis, E. Ultimate Waveform Reproducibility of Extreme-Ultraviolet Pulses by High-Harmonic Generation in Quartz. *Nat. Photonics* **2018**, *12*, 291–296. https://doi.org/10.1038/s41566-018-0123-6.
(27) Salières, P.; L'Huillier, A.; Lewenstein, M. Coherence Control of High-Order Harmonics. *Phys. Rev. Lett.* **1995**, *74*, 3776–3779. https://doi.org/10.1103/PhysRevLett.74.3776.
(28) Vampa, G.; Mcdonald, C. R.; Orlando, G.; Corkum, P. B.; Brabec, T. Semiclassical Analysis of High Harmonic Generation in Bulk Crystals. *Phys Rev. B* **2015**, *91*, 064302. https://doi.org/10.1103/PhysRevB.91.064302.
(29) Du, T. Y.; Tang, D.; Huang, X. H.; Bian, X. Bin. Multichannel High-Order Harmonic Generation from Solids. *Phys. Rev. A* **2018**, *97*, 1–7. https://doi.org/10.1103/PhysRevA.97.043413.
(30) Lu, J.; Cunningham, E. F.; You, Y. S.; Reis, D. A.; Ghimire, S. Interferometry of Dipole Phase in High

Harmonics from Solids. *Nat. Photonics* **2019**, *13*. https://doi.org/10.1038/s41566-018-0326-x.
(31) Wu, M.; Ghimire, S.; Reis, D. A.; Schafer, K. J.; Gaarde, M. B. High-Harmonic Generation from Bloch Electrons in Solids. *Phys. Rev. A - At. Mol. Opt. Phys.* **2015**, *91*, 1–11. https://doi.org/10.1103/PhysRevA.91.043839.
(32) Tamaya, T.; Ishikawa, A.; Ogawa, T.; Tanaka, K. Diabatic Mechanisms of Higher-Order Harmonic Generation in Solid-State Materials under High-Intensity Electric Fields. *Phys. Rev. Lett.* **2016**, *116*, 1–5. https://doi.org/10.1103/PhysRevLett.116.016601.
(33) Palik, E. D. *Handbook of Optical Constants of Solids*; Academic Press, 1998.

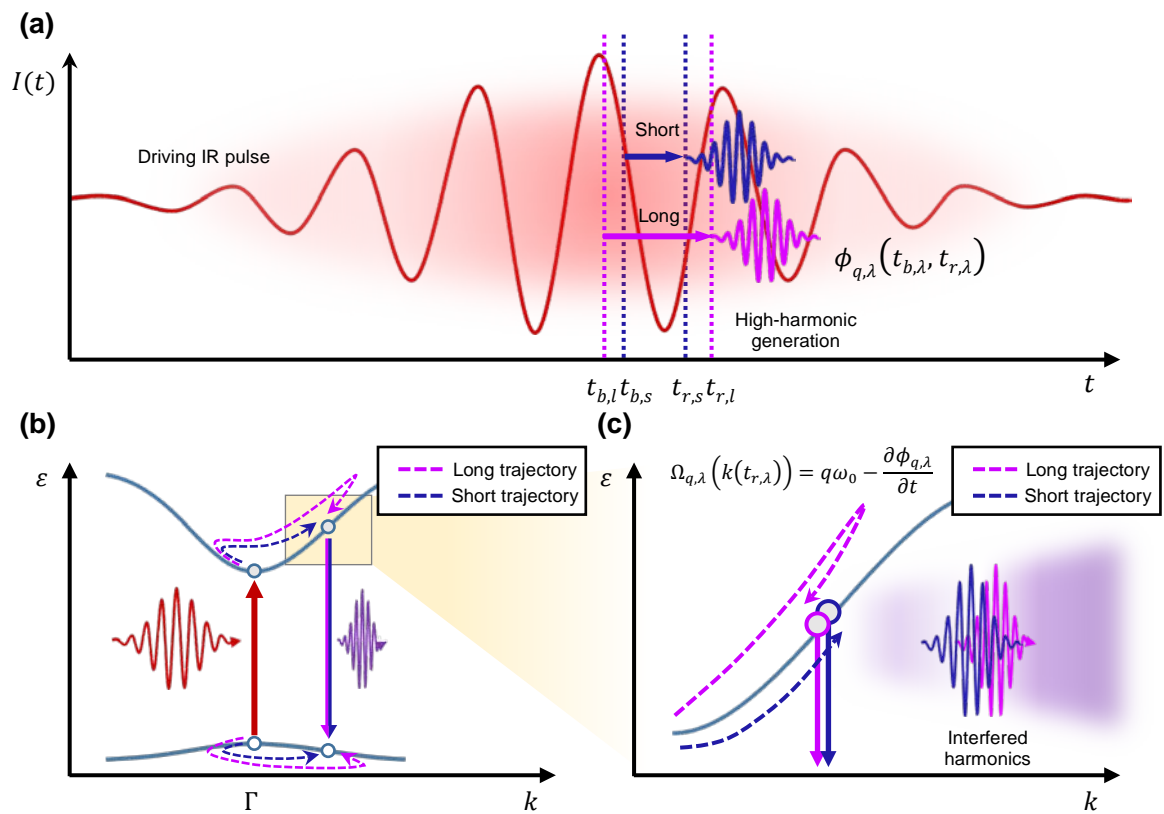

**Figure 1.** Long and short trajectories in solids. (a) Long ($l$) and short ($s$) trajectories depicted relative to the driving laser field in the temporal domain. $t_b$ ($t_r$) denotes the time of birth (recombination) of the electron-hole pair. $\phi_{q,\lambda}$ is the accumulated phase of the $q$-th harmonic with $\lambda$ being either $l$ or $s$ trajectory. (b) Long and short trajectories traced within a crystal band structure. (c) Enlarged view of the recombination process followed by interference between the long and short trajectories. $\Omega_{q,\lambda}$ is the chirped center frequency of the $q$-th harmonic.

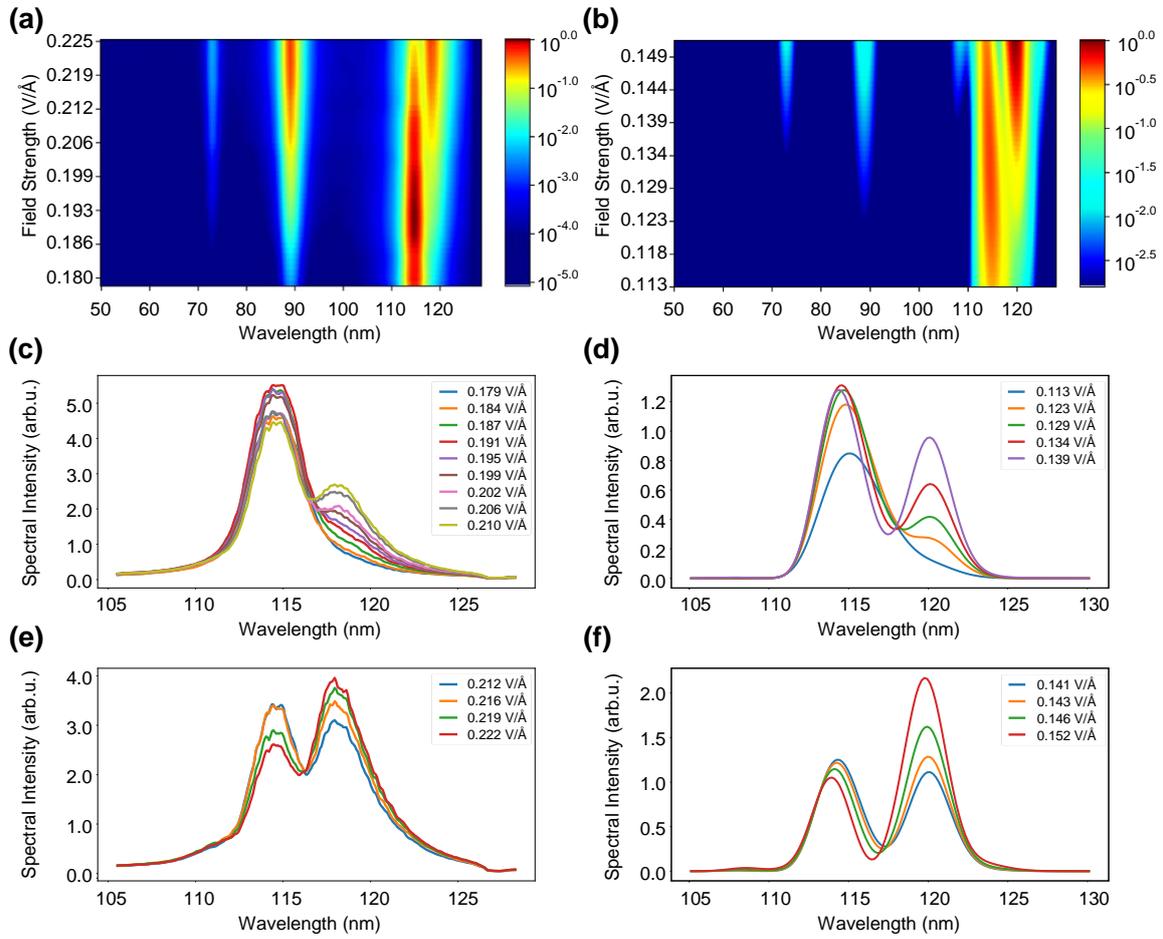

**Figure 2.** Harmonic spectrum vs. field strength. (a) (Experiment) harmonic spectra with varying field strength. (b) (Simulation) calculated spectra with varying field strength. (c) (Experiment) spectral variation of the 7-th harmonic for relatively low field strengths. (d) (Simulation) calculated spectral variation of the 7-th harmonic for low field strengths. (e) (Experiment) peak reversal of the 7-th harmonic for high field strengths. (f) (Simulation) peak reversal of the 7-th harmonic for high field strengths.

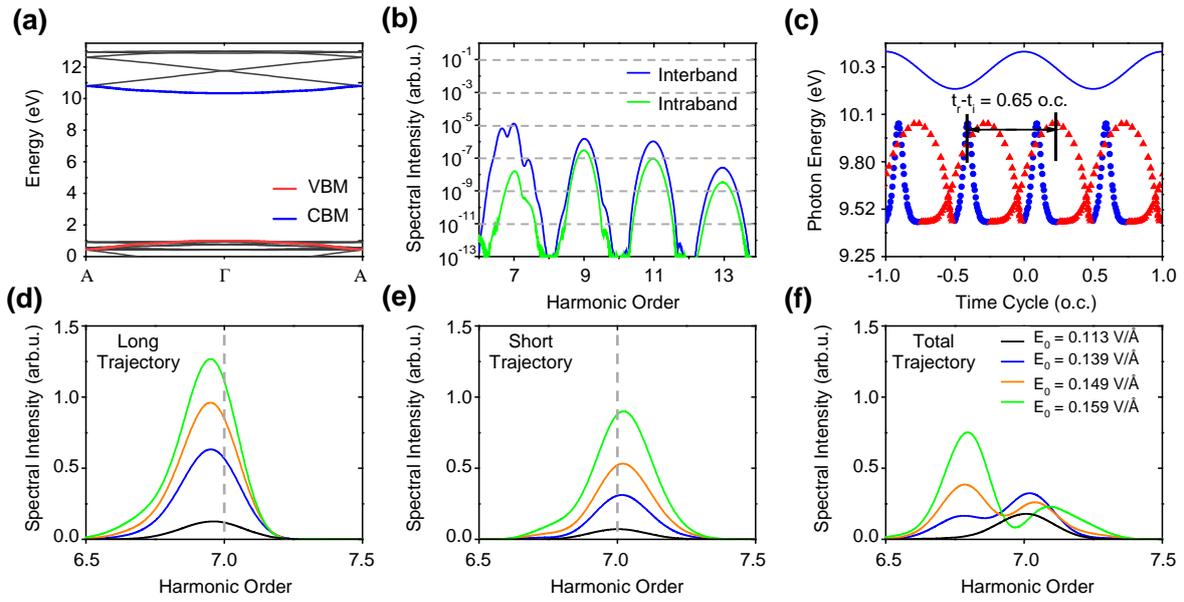

**Figure 3.** Long and short trajectory contributions. (a) Band structure of sapphire along the Γ-A direction. VBM and CBM are the valence and conduction band models finally selected for simulation. (b) Expected interband and intraband contributions ($E_0$=0.139 V/Å). (c) Harmonic photon energy as a function of ionization (blue line) and recombination (red line) times. Shown at top is the driving laser oscillation cycle (o.c.). The excursion time of the long trajectory is given as 0.65 o.c. between the birth time ($t_b$) and recombination time ($t_r$). (d-f) Calculated harmonic spectra with varying laser intensity for long, short and coherently combined trajectories. The dephasing time is given as $T_2$=2.67 fs (1.0 o.c.).

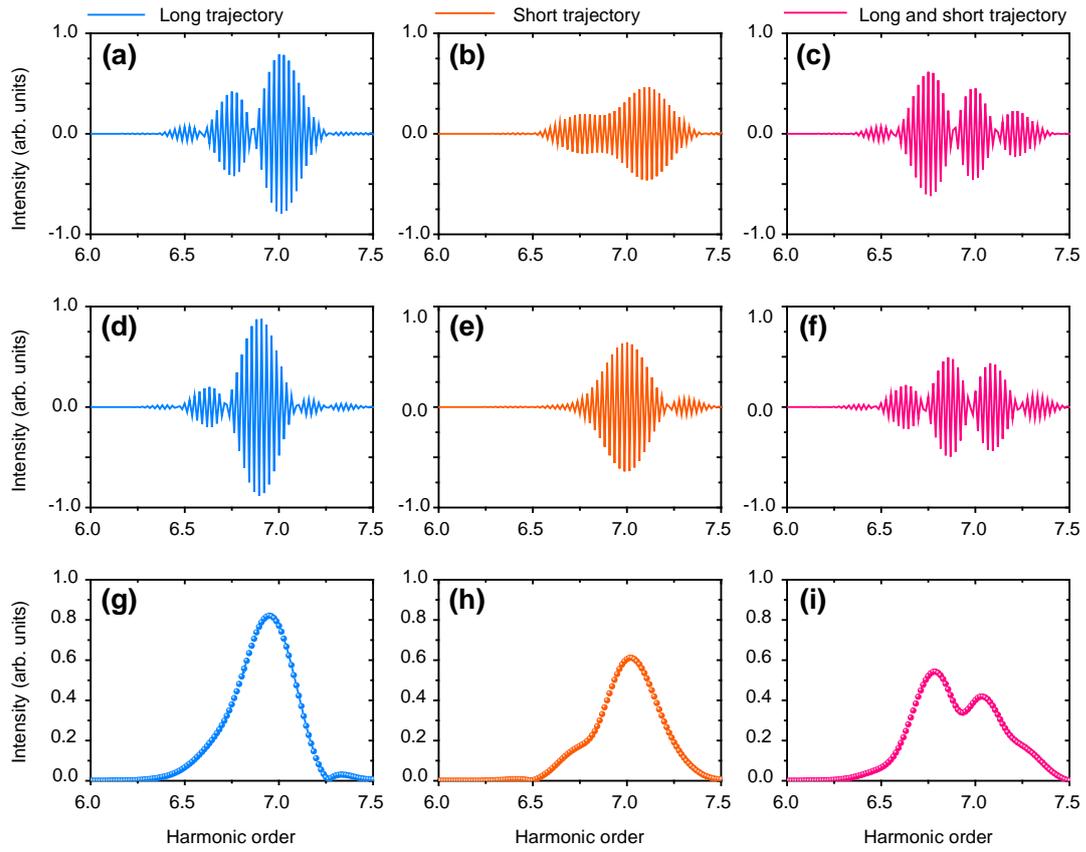

**Figure 4.** Interference of long and short trajectories. (a-c) Real parts of interband polarization for long, short and interfered trajectories. (d-f) Imaginary parts of interband polarization for long, short and interfered trajectories. (g-h) Absolute amplitudes of interband polarization for long, short and interfered trajectories. The dephasing time is given as $T_2$=2.67 fs (1.0 o.c.) for a field strength ($E_0$=0.149 V/Å).

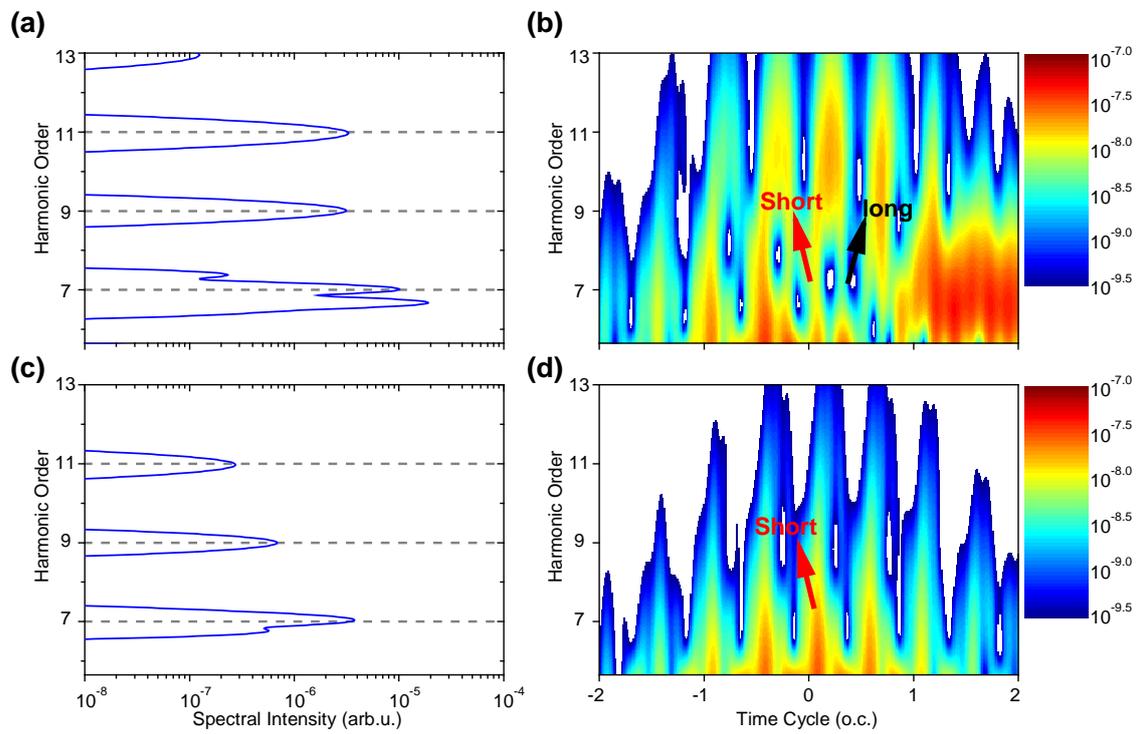

**Figure 5.** Dephasing time effect on harmonic spectral modulation. (a),(b) High harmonic spectrum and spectrograph calculated for a dephasing time of $T_2$=2.67 fs (1.0 o.c.). (c),(d) High harmonic spectrum and spectrograph calculated for a dephasing time of $T_2$=0.668 fs (0.25 o.c.). The field strength is set at $E_0$=0.149 V/Å for all plots.

*Supporting Information*

# Spectral Interference of High Harmonic Generation in Solids

Yong Woo Kim[1], Tian-Jiao Shao[2,3], Hyunwoong Kim[1], Seunghwoi Han[4], Seungchul Kim[5], Marcelo Ciappina[6], Xue-Bin Bian[2], Seung-Woo Kim[1]

[1]Department of Mechanical Engineering, Korea Advanced Institute of Science and Technology (KAIST), 291 Daehak-ro, Yuseong-gu, Daejeon 34141, South Korea

[2]State Key Laboratory of Magnetic Resonance and Atomic and Molecular Physics, Wuhan Institute of Physics and Mathematics, Chinese Academy of Sciences, Wuhan 430071, China

[3] School of Physical Sciences, University of Chinese Academy of Sciences, Beijing 100049, China

[4]Institute for the Frontier of Attosecond Science and Technology, CREOL and Department of Physics, University of Central Florida, Orlando, Florida 32816, USA

[5]Department of Optics and Mechatronics Engineering, College of Nanoscience and Nanotechnology, Pusan National University, 2 Busandaehak-ro 63beon-gil, Busan 46241, South Korea

[6]Institute of Physics of the ASCR, ELI-Beamlines project, Na Slovance 2, 182 21 Prague, Czech Republic

**Experimental setup of high harmonic generation from sapphire.**

Figure S1 shown below illustrates the schematic of the apparatus used to obtain the experimental data given in Fig. 2 of the main text.

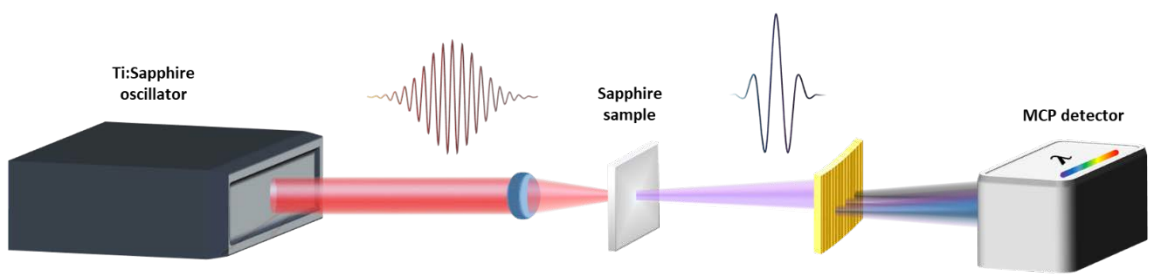

**Figure S1. Experimental setup.** The driving NIR pulses produced from a Ti:sapphire oscillator is focused on the sapphire sample. The high harmonics spectrum is detected through a diffraction grating using a micro-channel plate (MCP).

**Theoretical model using semiconductor Bloch equations**

The electron and hole dynamics within a solid crystal is governed by the semiconductor Bloch equations (SBEs) in the reciprocal momentum space, which is written in a two-band model as [1, 2],

$$\dot{\pi}(K,t) = -\frac{\pi(K,t)}{T_d} - i\Omega(K,t)w(K,t)e^{-iS(K,t)} \quad (1.1)$$

$$\dot{n}_b(K,t) = is_b\Omega^*(K,t)\pi(K,t)e^{iS(K,t)} + c.c. \quad (1.2)$$

In the above equations, $\pi(K,t)$ denotes the polarization strength between the conduction (C) band and the valence (V) band, and $n_b(K,t)$ represents the charge carrier population within the conduction $(b \to C)$ and valence $(b \to V)$ bands. Besides, the Rabi frequency is given as $\Omega(K,t) = d(K + A(t))E(t)$, the population inversion as $w(K,t) = n_C(K,t) - n_V(K,t)$, and the distinguishing factor as $s_b = 1$ for C band and -1 for V band. Furthermore, $S(K,t) = \int_{-\infty}^{t} \varepsilon_g(K + A(t'))dt'$ is the classical action term with $\varepsilon_g$ being the band gap energy, and $T_d$ is the dephasing time related to the coherence between the electron-hole pair in the C and V bands.

The SBEs can be solved to obtain the intraband current ($j_{ra}$) and interband current ($j_{er}$) as

$$j_{ra}(t) = \sum_{b=c,v} \int_{\overline{BZ}} v_b[K + A(t)]n_b(K,t)d^3K \quad (2.1)$$

$$j_{er}(t) = \frac{d}{dt}\int_{\overline{BZ}} p(K,t)d^3K \quad (2.2)$$

where $v_b[K + A(t)]$ denotes the group velocity and $p(K,t) = d(K + A(t))\pi(K,t)e^{iS(K,t)} + c.c.$ is the interband polarization. The two current quantities are strongly dependent on the curvature of the band structure as well as the transition dipoles. Note that the above integrations are carried over the whole Brillouin zone ($\overline{BZ}$), being restricted to the one-dimensional path along the driving laser polarization in our calculation. The interband and intraband contributions to harmonics can be separated individually from the Fourier transform (FT) of $j_{er}(t)$ and $j_{ra}(t)$ as $|FT\{j_{er}(t)\}|^2$ and $|FT\{j_{ra}(t)\}|^2$. The combined contribution of $j_{tot}(t) = j_{er}(t) + j_{ra}(t)$ is obtained by $|FT\{j_{tot}(t)\}|^2$.

**Saddle-point analysis for semiclassical trajectory tracing**

By using the Keldysh approximation, $w(K,t) \approx 1$, the interband current can be written as [2],

$$j_{er}(t) = \frac{d}{dt}\int_{BZ} d^3K d(K)\left[\int_{-\infty}^{t} dt' E(t')d^*(K+A(t'))e^{-iS(K,t',t)-\frac{t-t'}{T_d}} + c.c.\right] \quad (3.1)$$

Note that the above approximation is appropriate, considering the ionization rate in solids is known to be usually less than a few percent. Then in the frequency domain, the harmonics generated from the interband current can be written as,

$$j_{er}(\omega) = \omega \int_{BZ} d^3K d(K) \int_{-\infty}^{\infty} dt\, e^{-i\omega t}\left[\int_{-\infty}^{t} dt' E(t')d^*(K+A(t'))e^{-iS(K,t',t)-\frac{t-t'}{T_d}} + c.c.\right] \quad (3.2)$$

in which the phase is expressed as,

$$i\phi = -iS(K,t,t') - i\omega t - \frac{t-t'}{T_d} \quad (4)$$

Using the saddle-point method[2], the results are given as

$$\nabla_k \phi = \int_{t'}^{t} \Delta v\left(K + A(t'')\right)dt'' = 0 \quad (5.1)$$

$$\frac{d\phi}{dt'} = \varepsilon_g[K+A(t')] - \frac{i}{T_d} = 0 \quad (5.2)$$

$$\frac{d\phi}{dt} = \varepsilon_g[K+A(t)] - \omega + \frac{i}{T_d} = 0 \quad (5.3)$$

where $\Delta v(K) = v_c(K) - v_v(K)$ means the difference between the band velocities in C and V bands. The classical trajectories of the interband transition can be traced by solving Eq. (5.1) to Eq. (5.3) for $T_d = \infty$ so as to obtain the ionization time $t_i$, recombination time $t_r$ and emitted photon energy $\omega$.

**First principle calculation**

The α-$Al_2O_3$ hexagonal conventional cell has 30 atoms and belongs to the space group $R\bar{3}C$. The exchange-correlation functional is obtained using the local density approximation (LDA). The projector augmented wave (PAW) method is used to describe the ion-electron interactions with a plane-wave cutoff of 600 eV. The Brillouin zone of the unit cell is sampled by a 6×6×2 k-point mesh in our self-consistent converged calculation. The optimized lattice parameters for α-Al$_2$O$_3$ (a=4.697 Å, c=12.793 Å) are in good agreement with reported experimental data (trigonal structure: a=4.76Å, c=12.99 Å) [3]. A bandgap value of 9.45 eV is used here, which is adopted from experimental measurements [4, 5].

**Extension of the spectral modulation phenomena to higher harmonics**

Figure S2 shows an extra simulation result made on the 9$^{th}$ harmonic with increasing the driving laser field strength. For relatively lower laser fields, as in the 7$^{th}$ harmonic discussed in the main text, the intensity-dependent phenomenon of spectral splitting and broadening is not significant, but it gradually develops through a complete peak reversal as the laser field increases. This simulation result justifies the possibility of extending the quantum path interference in harmonics generation to shorter wavelengths, suggesting that at extreme intensities, spectral modulation and control can be made for arbitrary harmonics, even in the far EUV and X-ray regime.

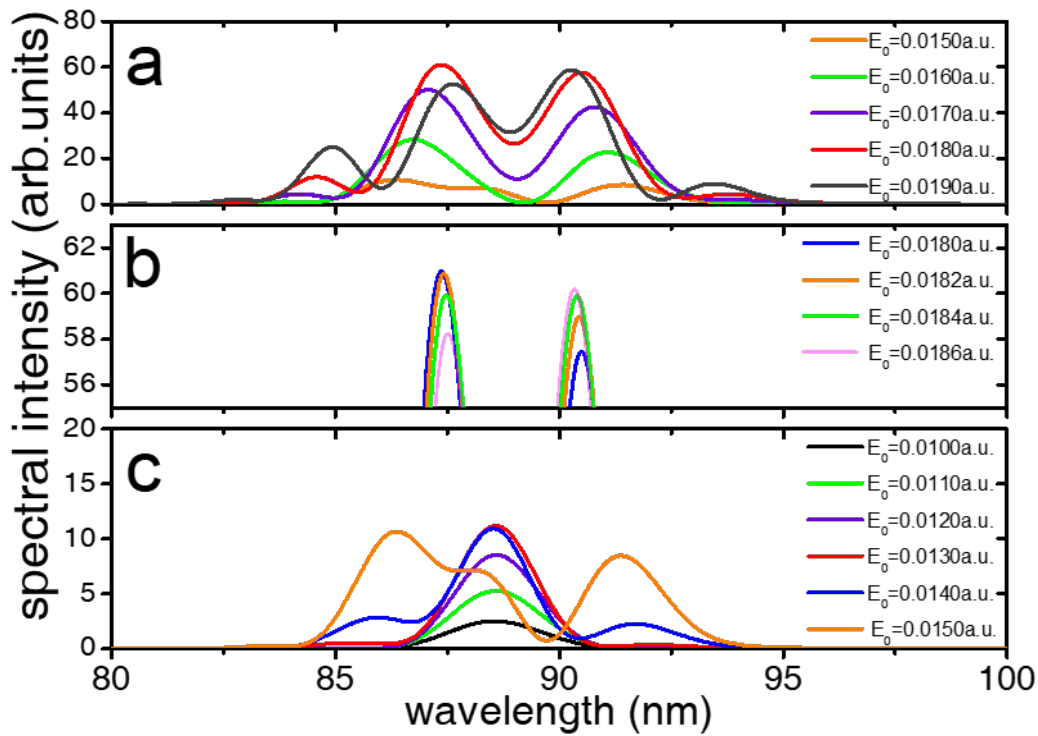

**Figure S2. Theoretical prediction for splitting of the 9$^{th}$ order harmonic under high driving laser field strength.** (a) Calculated 9$^{th}$ harmonic spectra for high laser fields between E$_0$=0.015 a.u. and E$_0$=0.019 a.u.; (b) Zoomed plot exhibiting a peak reversal for field strength between E$_0$=0.0180 a.u. and E$_0$=0.0186 a.u.; (c)

Calculated 9th harmonic spectra for lower laser fields between $E_0=0.010$ a.u. and $E_0=0.015$ a.u. All data are obtained by solving the SBEs with a dephasing time of $T_2=2.67$ fs (1.0 o.c.).

**REFERENCES**

(1) Vampa, G.; McDonald, C. R.; Orlando, G.; Klug, D. D.; Corkum, P. B.; Brabec, T. Theoretical Analysis of High-Harmonic Generation in Solids. *Phys. Rev. Lett.* **2014**, *113* (7), 073901.

(2) Vampa, G.; Mcdonald, C. R.; Orlando, G.; Corkum, P. B.; Brabec, T. Semiclassical Analysis of High Harmonic Generation in Bulk Crystals. *Phys Rev. B* **2015**, *91* (6), 064302.

(3) Aldebert, P.; Traverse, J. P. Neutron Diffraction Study of Structural Characteristics and Ionic Mobility of α - Al2O3 at High Temperatures. *J. Am. Ceram. Soc.* **1982**, 65 (9), 460.

(4) French, R. H.; Jones, D. J.; Loughin, S. Interband Electronic Structure of α-Alumina up to 2167 K. *J. Am. Ceram. Soc.* **1994**, 77 (2), 412.

(5) Marinopoulos, A. G.; Grüning, M. Local-field and excitonic effects in the optical response of α-alumina. *Phys Rev B* **2011**, 83 (19), 195129.